\documentclass[epj]{svjour}

\usepackage{graphicx}
\usepackage{fancyhdr}
\def\makeheadbox{{
 }}
\newcommand{\MET}{\mbox{$\not\hspace{-0.11cm}E_T$}}

\def\mytitle{My title} 
\def\myauthors{My name}  
\def\mytype{My type of session}
\def\mysession{My session}

\def\mytitle{Searches for Extra Dimensions and for Heavy Resonances with the D{\O} Detector}
\def\myauthors{Carsten Magass}
\def\mytype{Contributed Talk}    
\def\mysession{Alternatives}

\begin{document}
\title{Searches for Extra Dimensions and for Heavy Resonances in 
Dilepton, Diphoton, Electron + Photon and Electron + Missing $\bf{E_T}$ 
Final States with the D{\O} Detector}

\author{Carsten Magass on behalf of the D{\O} Collaboration
\thanks{\emph{Email:} magass@fnal.gov}}
\institute{III. Physikalisches Institut A, RWTH Aachen, D-52056 Aachen (Germany)}
\date{}
\abstract{
The high mass spectrum of lepton and photon pairs is sensitive to a broad 
array of new physics. Examples include searches for extra dimensions in the 
dielectron and diphoton channels. A direct search for electron compositeness 
is possible in the production of excited electrons decaying into an electron 
and a photon. In addition, the electron plus missing transverse energy data 
sample can be searched for a $W'$ boson. Latest results in searches in the 
high mass dielectron, diphoton, electron plus photon, and electron plus 
missing transverse energy channels obtained by the D{\O} experiment at 
the Tevatron are reported, using a data set corresponding to an integrated 
luminosity of about 1~fb$^{-1}$. Since no significant excess is
observed in the data in all cases, limits are set which improve on 
previous searches.
} 
\maketitle
%

\section{Introduction}

The standard model (SM) describes the fundamental fermions and their 
interactions via gauge bosons at a high level of accuracy, but it is 
not considered to be a complete theory. Some aspects -- like the 
number of fermion families and the hierarchy problem -- remain 
unexplained. Many possible extensions to the SM have been proposed, 
which usually predict new particles and contain new parameters. 
These new particles and their decays can be looked for at 
a collider experiment like the proton-antiproton collider Tevatron 
which currently (Run II) operates at a center-of-mass energy 
of $\sqrt{s}=1.96$~TeV, thus being the world's most energetic 
particle collider until the startup of the Large Hadron Collider (LHC).
At these colliders, it is possible to produce particles with extremely 
high masses that could never be explored before.

The Tevatron collider is performing very well; an integrated luminosity 
of more than 3~fb$^{-1}$ has already been delivered. Three recent 
analyses \cite{d0results} are presented which use data taken with 
the D{\O} detector \cite{d0det} until February 2006 (Run IIa). This dataset 
corresponds to a recorded luminosity of about 1~fb$^{-1}$.
The data are searched for new particles introduced in different extensions 
to the SM via their decays into high energetic electrons, photons and 
neutrinos, where the latter ones give rise to missing transverse 
energy ($\MET$). The analyses make use of the capabilities to detect 
and identify such particles with the well understood D{\O} detector.

\section{Randall-Sundrum Extra Dimensions}

\begin{figure*}
\includegraphics[width=0.276\textwidth,angle=90]{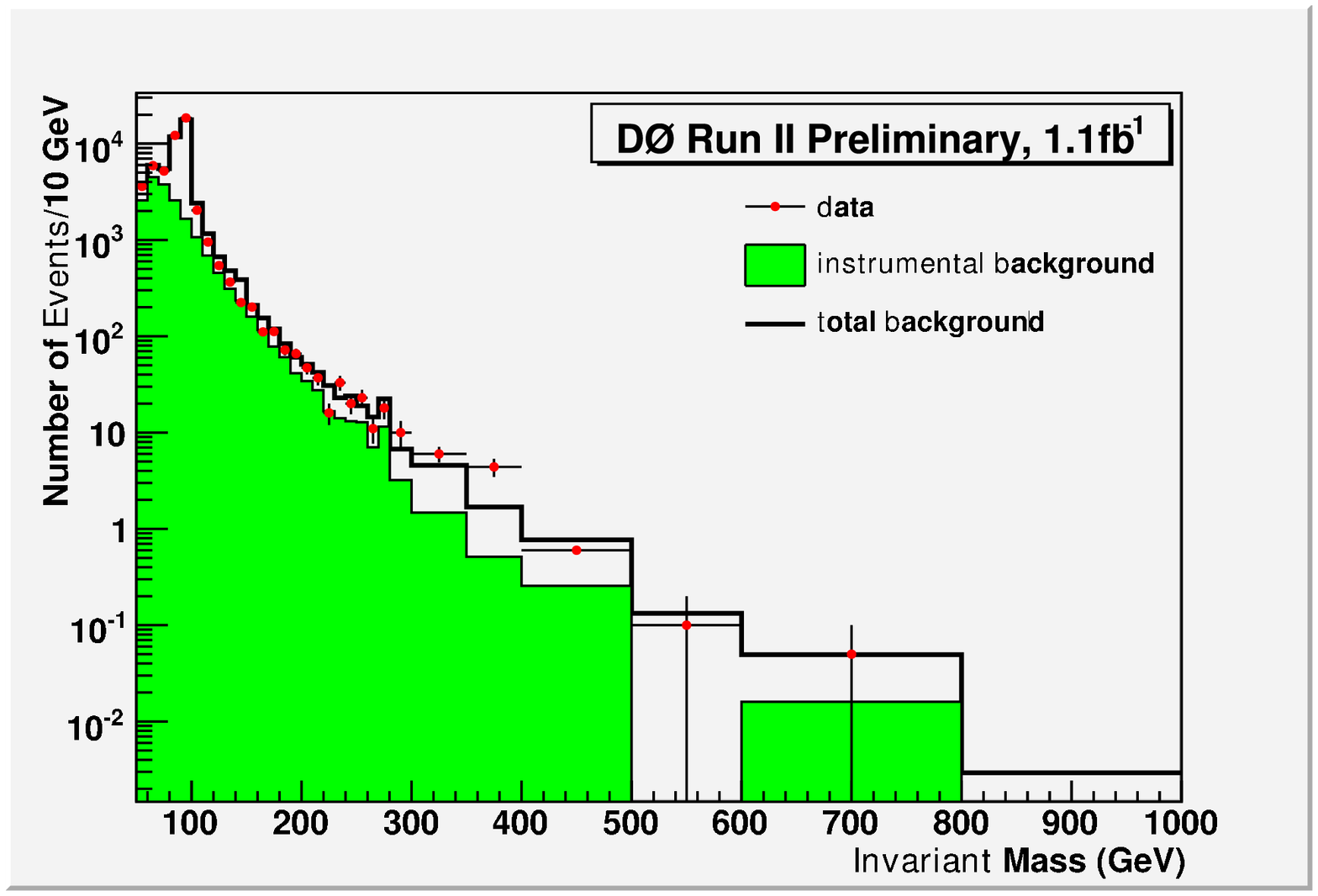} \hfill
\includegraphics[width=0.29\textwidth,angle=0]{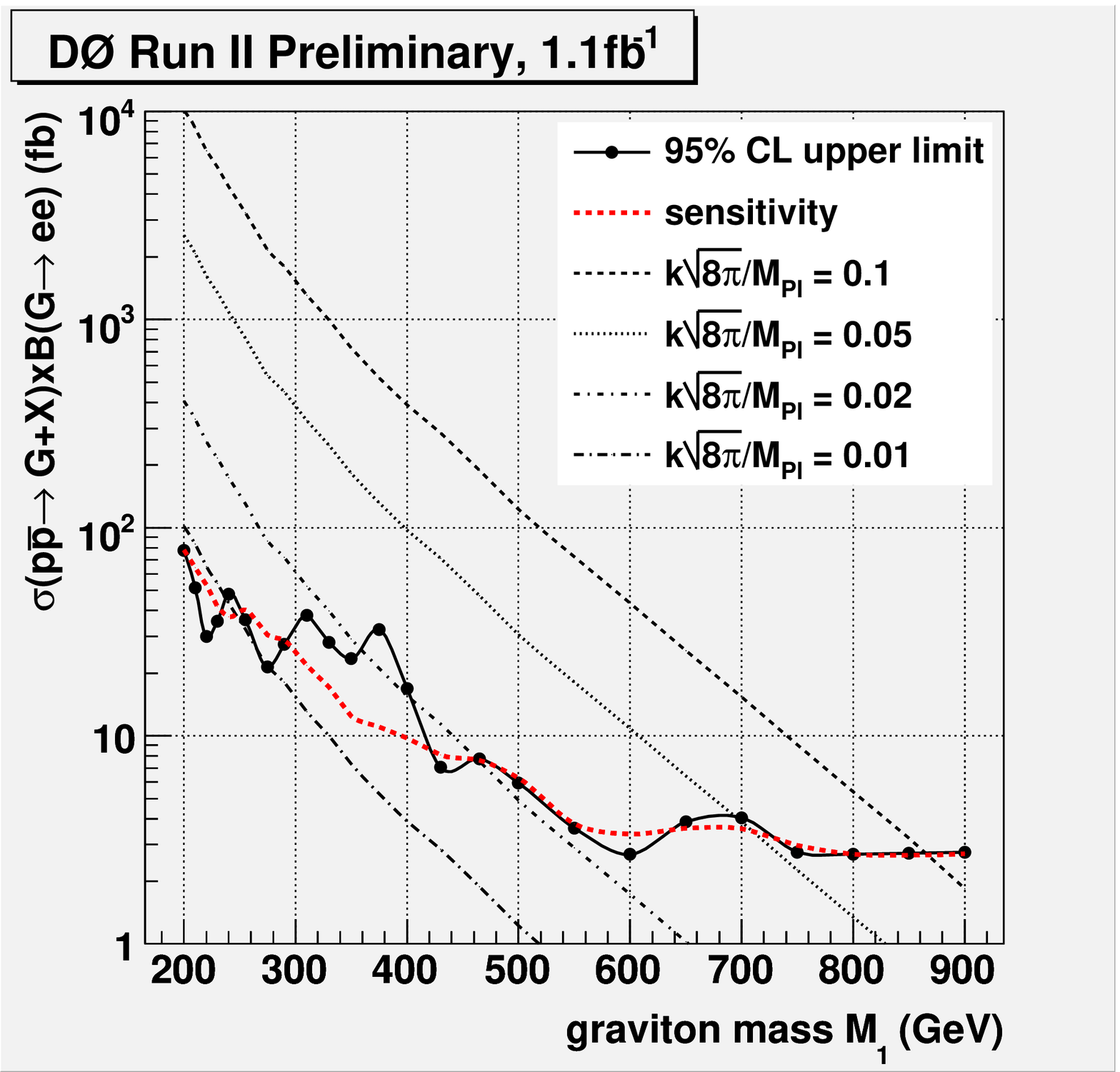} \hfill
\includegraphics[width=0.29\textwidth,angle=0]{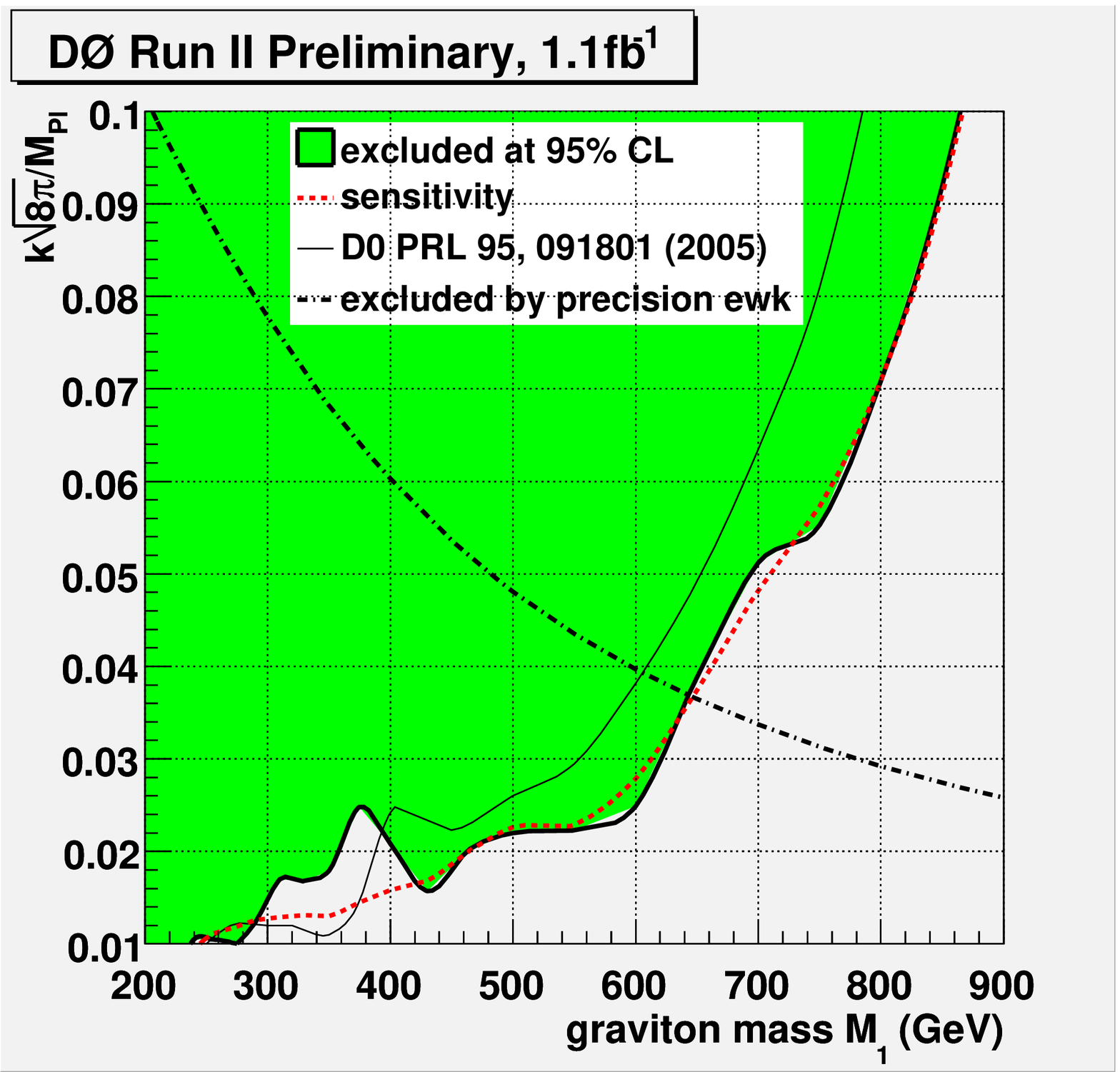}
\caption{Left: invariant mass spectrum of the two EM objects; 
middle: 95\% CL upper limit on the production cross section times 
branching fraction compared with the sensitivity and theoretical 
prediction for various values of the coupling; right: 95\% CL exclusion 
contour in the ($k\sqrt{8\pi}/M_{\rm Planck},M_1$)-plane compared with 
the sensitivity and the previously published contour.}
\label{fig:RS}  
\end{figure*}

The hierarchy problem posed by the large difference between the Planck 
scale $M_{\rm Planck} \approx 10^{16}$~TeV, at which gravity is expected 
to become strong, and the scale of the electroweak symmetry breaking 
at $\approx 1$~TeV can be solved by the introduction of extra spatial 
dimensions. In the model of Randall and Sundrum \cite{RS} the gravity 
originating on a (3+1)-dimensional brane (Planck brane) is separated 
from the SM brane in a 5$^{th}$ dimension with a warped metric. 
Gravity appears weak at the SM brane due to the exponential suppression 
caused by the metric. In the simplest version of this model the only 
particles that can propagate in the extra dimension are gravitons, 
which appear as towers of Kaluza-Klein excitations with masses and widths
that are determined by the parameters of the model: the mass of the first 
excited mode of the graviton, $M_1$, and a dimensionless coupling to 
SM fields, $k\sqrt{8\pi}/M_{\rm Planck}$. Precision electroweak data 
and the requirement that the model remains perturbative constrain the 
coupling to lie between about 0.01 and 0.1.

The resonant production of the first excited graviton mode and its decay 
into dielectron and diphoton pairs are investigated, 
$p\bar p\rightarrow G +X\rightarrow ee/\gamma\gamma+X$. The dielectron 
pairs from the decaying graviton with spin 2 are in a $p$-wave state, 
whereas the $\gamma\gamma$ pairs can also be in an $s$-wave state. 
Because of this the branching fraction into $\gamma\gamma$ is
twice the branching fraction into $ee$.

Data events are selected if they are triggered and contain two isolated 
objects in the central electromagnetic (EM) calorimeter ($|\eta|<1.1$) 
with transverse energy $E_T>25$~GeV. Further, the energy deposition 
patterns of the two objects are required to be consistent with EM showers. 
A track match criterion is omitted in order to accept both $\gamma\gamma$ 
and $ee$ decay channels. The dataset contains 50,354 events with an 
invariant mass of the two EM objects above 50~GeV. 

Physics backgrounds (Drell-Yan production of $ee$ and direct $\gamma\gamma$ 
production) are simulated using \textsc{pythia} \cite{pythia} Monte Carlo 
samples. Differences in reconstruction efficiencies observed in data and
Monte Carlo are taken into account. Instrumental backgrounds, in which 
one or both of the EM objects are misidentified, are estimated from data. 
For this purpose a dataset is used which contains two EM objects that fail 
the tight EM shower criteria. This provides an estimate for the shape of 
the invariant mass spectrum of the instrumental background. In order to
determine the relative contribution of this background, the invariant 
mass spectrum is fitted around the $Z$ peak (60~GeV~$<m_{ee}<$ 140~GeV) 
with the sum of physics and instrumental backgrounds, 
see Fig. \ref{fig:RS} (left).

The spectrum above 140~GeV is searched for a possible graviton signal. 
This is accomplished by constructing a sliding mass window, which is 
optimized with respect to the sensitivity for a given graviton mass. 
Systematic uncertainties on the signal and background samples are 
of the order of 10\%. 

Since no significant deviation from the background prediction is observed, 
upper limits on the production cross section times branching fraction, 
$\sigma(p\bar p\rightarrow G+X)\times B(G\rightarrow ee)$, are calculated 
using a Bayesian approach \cite{d0limit} with a flat prior. The resulting 
95\% confidence level (CL) limits (expected and observed) as a function of 
the graviton mass are shown in Fig. \ref{fig:RS} (middle) and compared to 
the theoretical predictions for various values of the coupling parameter. 
Fig. \ref{fig:RS} (right) displays the 95\% CL exclusion contour in the 
($k\sqrt{8\pi}/$ $M_{\rm Planck},M_1$)-plane. Hence, masses for the first 
excited graviton mode below 865~GeV (240~GeV) can be excluded at the 
95\% CL for $k\sqrt{8\pi}/M_{\rm Planck}=0.1$ (0.01).

\section{Excited Electrons}

\begin{figure*}
\includegraphics[width=0.32\textwidth,angle=0]{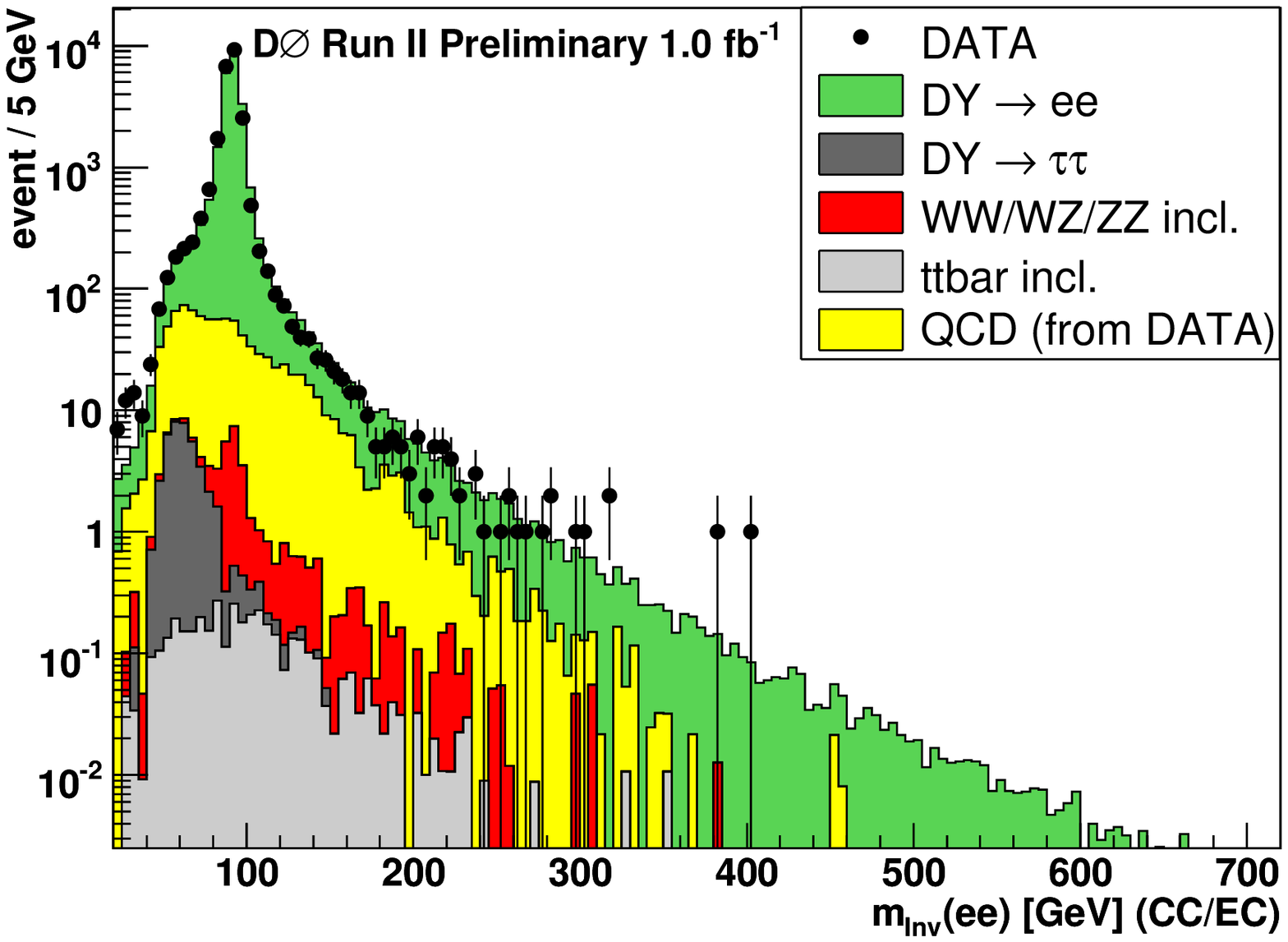} \hfill
\includegraphics[width=0.32\textwidth,angle=0]{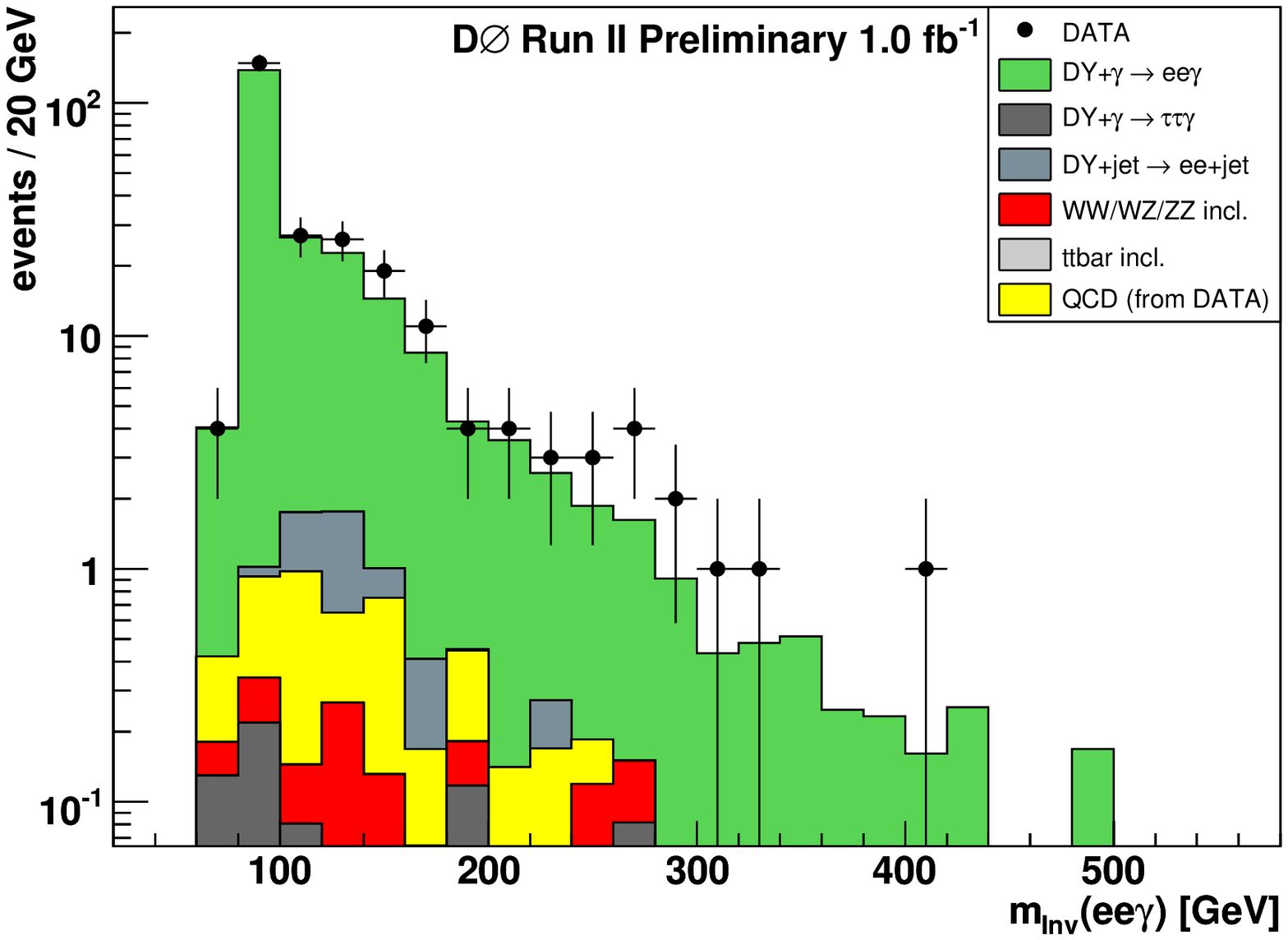} \hfill
\includegraphics[width=0.32\textwidth,angle=0]{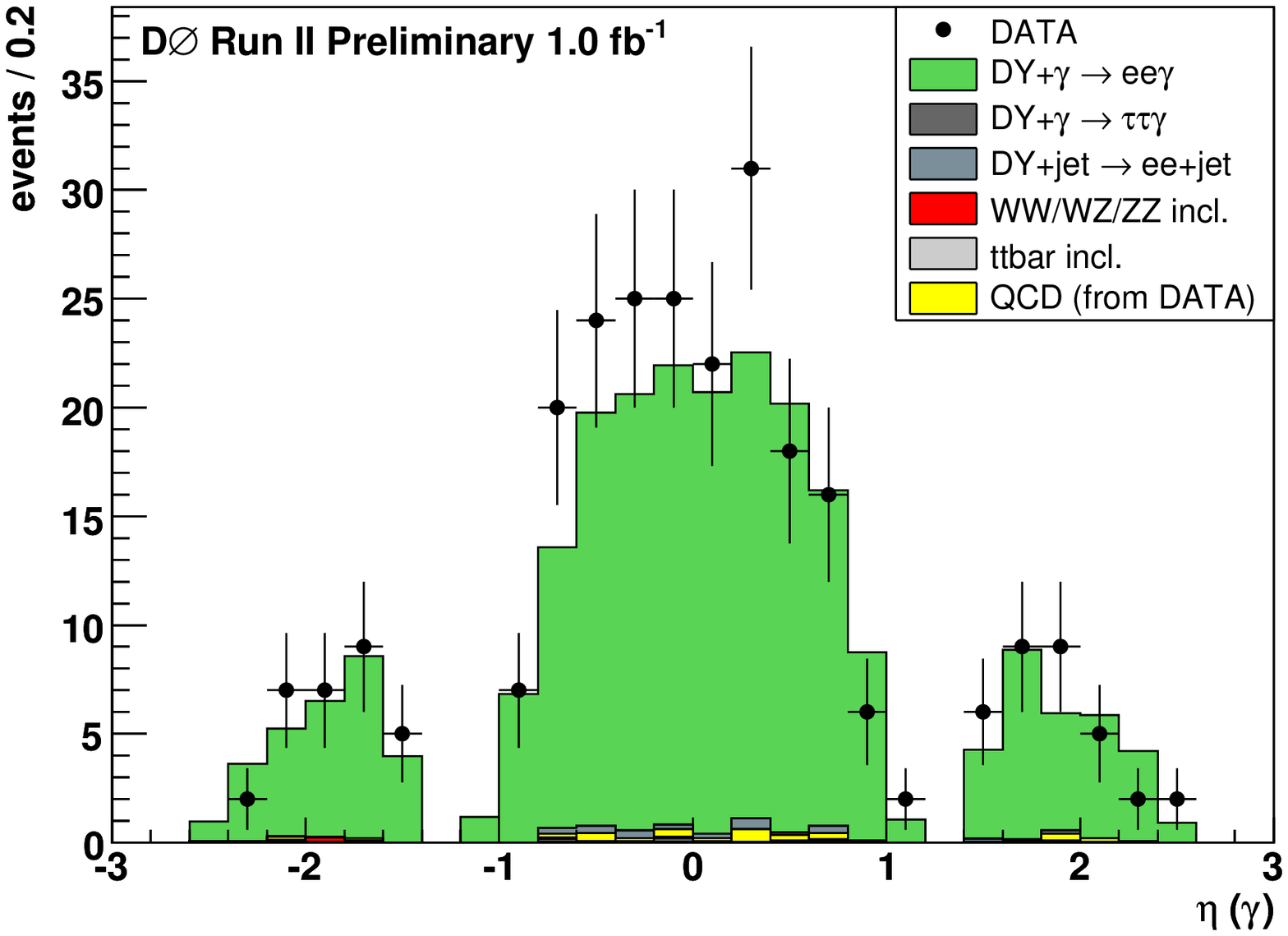}
\caption{Left: invariant mass distribution of the two electrons $m_{ee}$ 
(CC/EC combination); middle: invariant mass distribution of the two 
electrons and the photon $m_{ee\gamma}$; right: $\eta$ distribution 
of the photon.}
\label{fig:EE1}  
\end{figure*}

\begin{figure*}
\includegraphics[width=0.345\textwidth,angle=0]{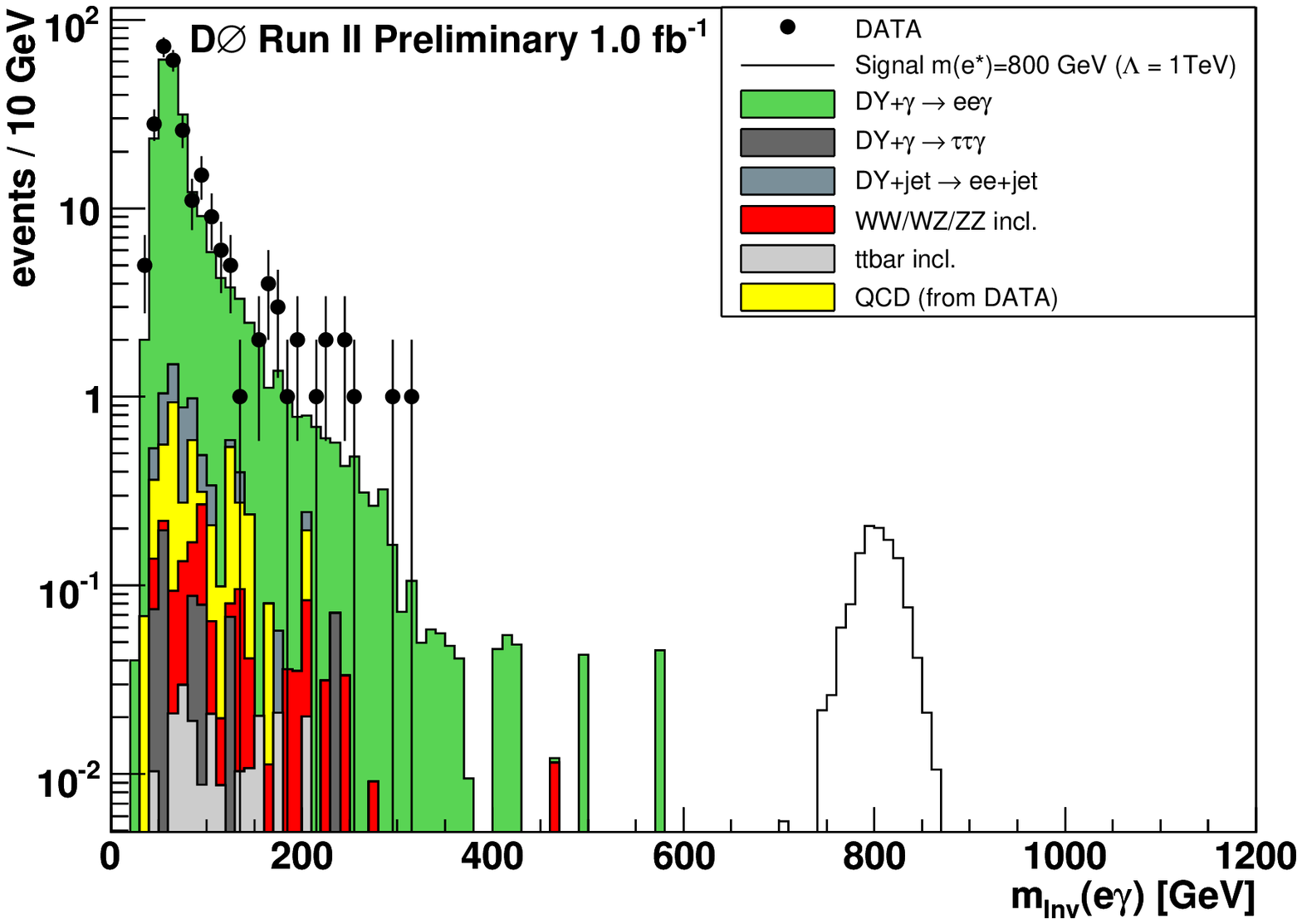} \hfill
\includegraphics[width=0.315\textwidth,angle=0]{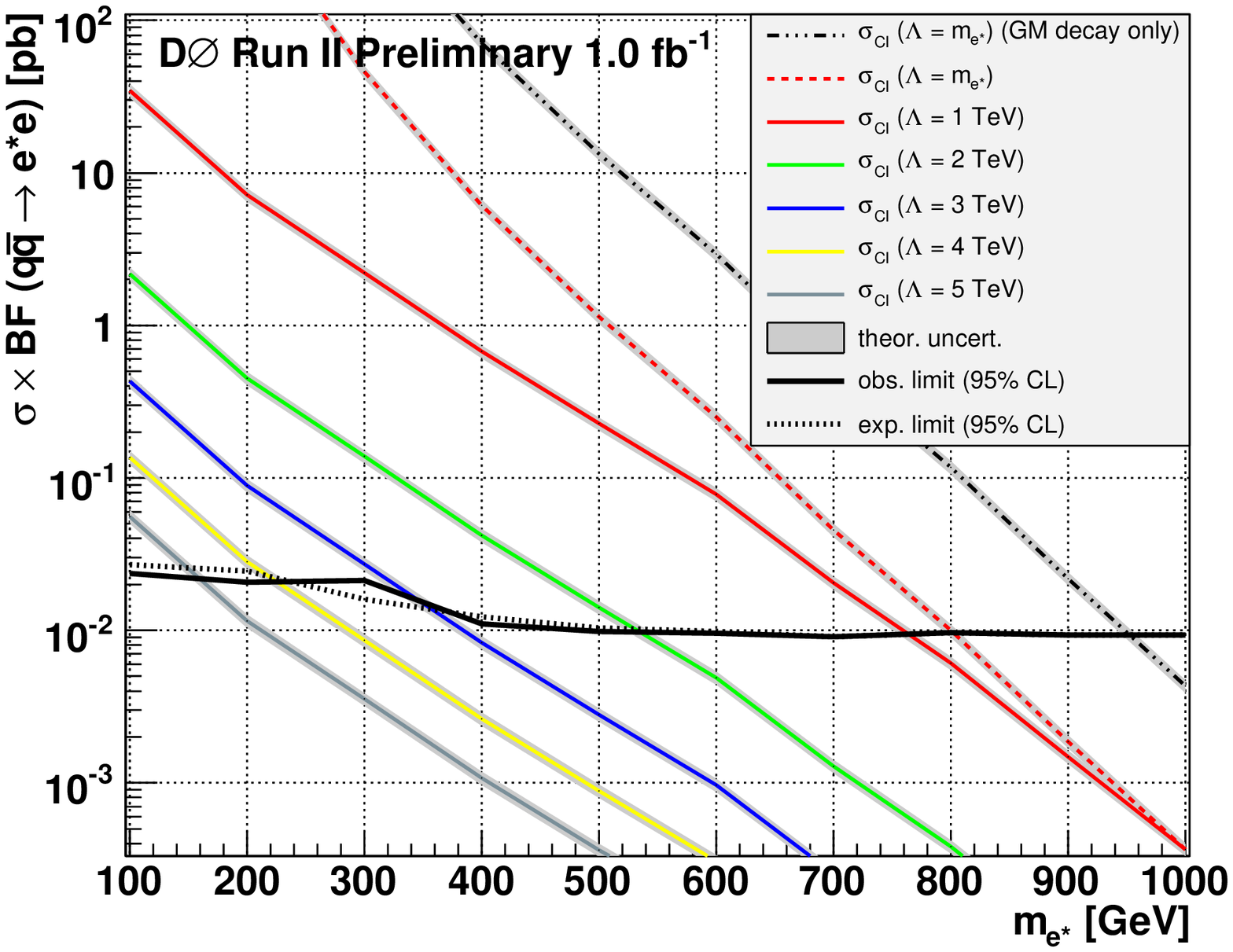} \hfill
\includegraphics[width=0.315\textwidth,angle=0]{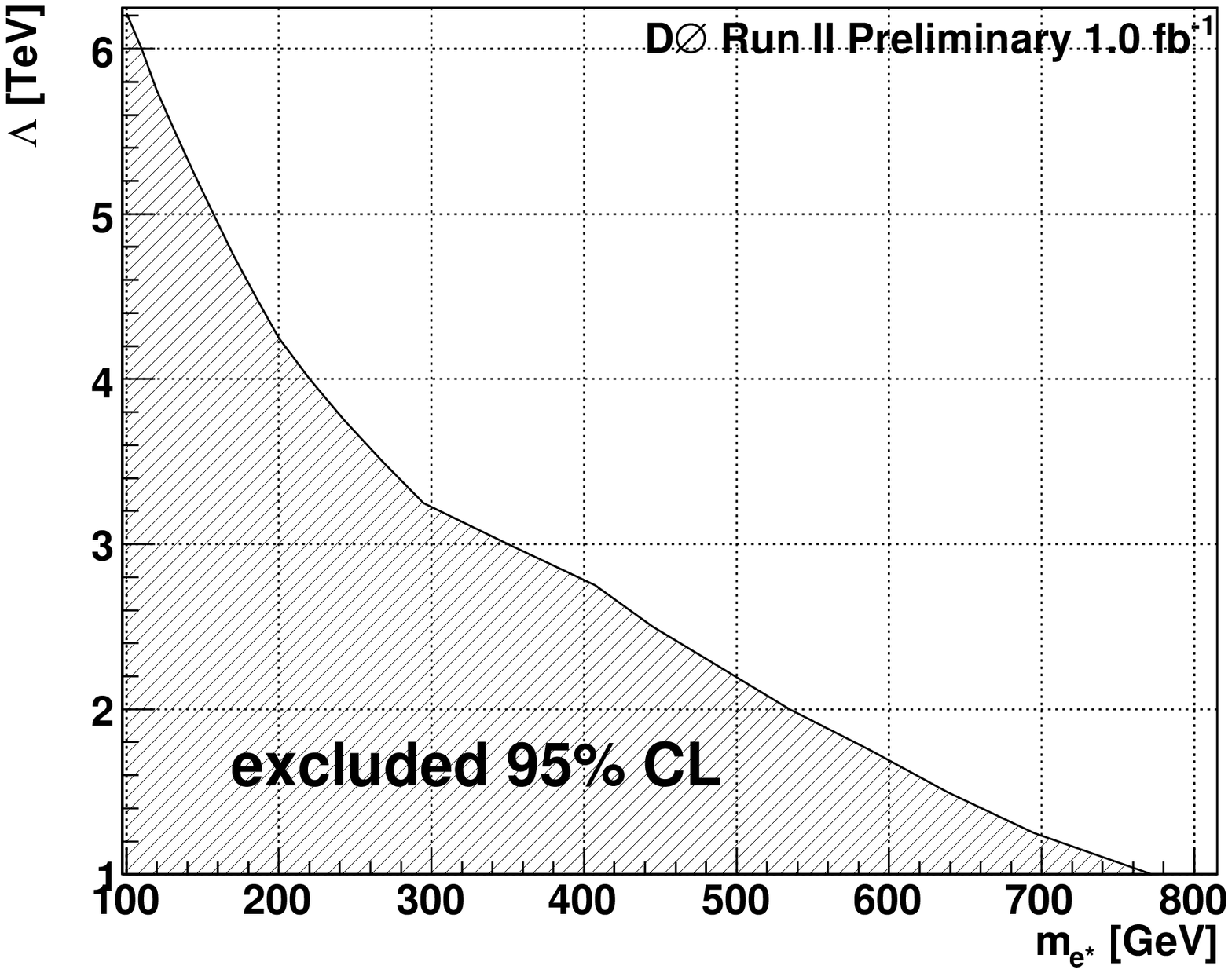}
\caption{Left: invariant mass distribution of the electron and 
photon (combination closest to $m_{e^*}=800$~GeV); middle: 95\% CL 
upper limits on the production cross section times branching fraction
as a function of the $e^*$ mass $m_{e^*}$; right: 95\% exclusion contour 
as a function of the compositeness scale $\Lambda$ and the excited 
electron mass $m_{e^*}$.}
\label{fig:EE2}  
\end{figure*}

In several models the observation of three families of fundamental 
fermions is explained by postulating that quarks and leptons are composed 
of scalar and spin-1/2 particles. In compositeness models \cite{compo} 
the underlying substructure leads to a large spectrum of excited states. 
In this analysis the production of excited electrons $e^*$ via
contact interaction (CI) and the subsequent electroweak decay 
$e^*\rightarrow e\gamma$ is investigated, 
$p\bar p \rightarrow ee^*+X \rightarrow ee\gamma+X$. The (gauge mediated) 
electroweak production of excited electrons is neglected due to its 
smallness compared to production via CI. Possible contributions from CI 
decays are taken into account in the decay width. Free parameters in 
this model are the compositeness scale $\Lambda$ and the mass of the 
excited electron $m_{e^*}$.

Data events are selected if they are triggered and contain two isolated 
electrons within $|\eta|<1.1$ (CC) or 1.5~$<|\eta|<$~2.5 (EC) with 
transverse energies $E_T>25$~GeV for the leading electron and 
$E_T>15$~GeV for the second leading electron. The electrons are required
to have energy deposition patterns consistent with EM showers, to be 
spatially separated, and to have an associated track. Combinations with 
both electrons found in opposite EC's are rejected.

This dataset is compared to the background prediction, consisting 
of \textsc{pythia} \cite{pythia} Monte Carlo samples for the physics 
backgrounds ($Z/\gamma^*\rightarrow ee$, 
$Z/\gamma^*\rightarrow \tau\tau\rightarrow eeX$ etc.) and a special 
sample for the instrumental background. The latter one is due to QCD 
multijet events with jets misidentified as electrons, and can be 
(similar to the Randall-Sundrum analysis) estimated from data. The 
normalization of the multijet sample is adjusted in the invariant 
mass interval 30~GeV~$<m_{ee}<$~65~GeV. The full dielectron data 
sample contains 62,930 events, which is compatible with the 
background expectation of 61,560~$\pm$~6,553 events, 
see Fig. \ref{fig:EE1} (left).

Finally, an additional photon within the CC or EC region of the 
calorimeter with $E_T>15$~GeV is selected. It is required to be 
separated from the two electrons, not to have a track associated, 
to be isolated and to fulfill tight EM shower constraints. 
The $ee\gamma$ dataset contains 259 events, while 232~$\pm$~29 
events are expected from background processes, see Fig. \ref{fig:EE1} 
(middle and right). Systematic uncertainties (backgrounds 10\%, signal 15\%) 
come from efficiency corrections, the luminosity, estimation of the 
photon misidentification rate, multijet estimation, cross sections and 
PDF uncertainties.

A crucial point is the reconstruction of the $e\gamma$ invariant mass, 
where the excited electron would be observable as a resonance. Depending 
on the mass of the excited electron two different approaches are considered:
For $m_{e^*}\leq 200$~GeV the lower energetic electron and the photon are 
combined; otherwise the combination closest to the searched $e^*$ mass is 
chosen. Fig. \ref{fig:EE2} (left) shows the $e\gamma$ invariant mass 
distribution for an excited electron with $m_{e^*}=800$~GeV. In order to 
further enhance the sensitivity for light excited electrons, combinations 
where both electrons and the photon are reconstructed in the EC are 
rejected (for $m_{e^*} \leq 300$~GeV), and a tighter cut on the spatial 
separation between the photon and the electron with lower energy is 
applied (for $m_{e^*} \leq 200$~GeV). 

Since no significant excess is observed in the data, upper limits are set 
on the CI production cross section times branching fraction, 
$\sigma(p\bar p\rightarrow ee^*+X) \times B(e^*\rightarrow e\gamma)$, 
using a Bayesian approach \cite{d0limit}. A final cut is applied in the 
$e\gamma$ invariant mass distribution, which is optimized with respect to 
the expected limit. The 95\% CL limits on the cross section as a function 
of the excited electron mass $m_{e^*}$ are shown in Fig. \ref{fig:EE2} (middle),
and compared to the theoretical prediction for various values of the 
compositeness scale $\Lambda$. The 95\% exclusion contour as a function of 
$\Lambda$ and $m_{e^*}$ is displayed in Fig. \ref{fig:EE2} (right). The lower 
mass limit for $\Lambda=1$~TeV is $m_{e^*}>756$~GeV, for $\Lambda=m_{e^*}$ 
the resulting lower mass bound is 796~GeV. If CI decays are neglected, a mass
limit of $m_{e^*}>946$~GeV is derived for $\Lambda=m_{e^*}$, thus improving 
the previous mass limit of $m_{e^*}>879$~GeV set by the CDF 
collaboration \cite{cdfestar}.

\section{New Heavy Charged Gauge Bosons}

\begin{figure*}
\includegraphics[width=0.31\textwidth,angle=0]{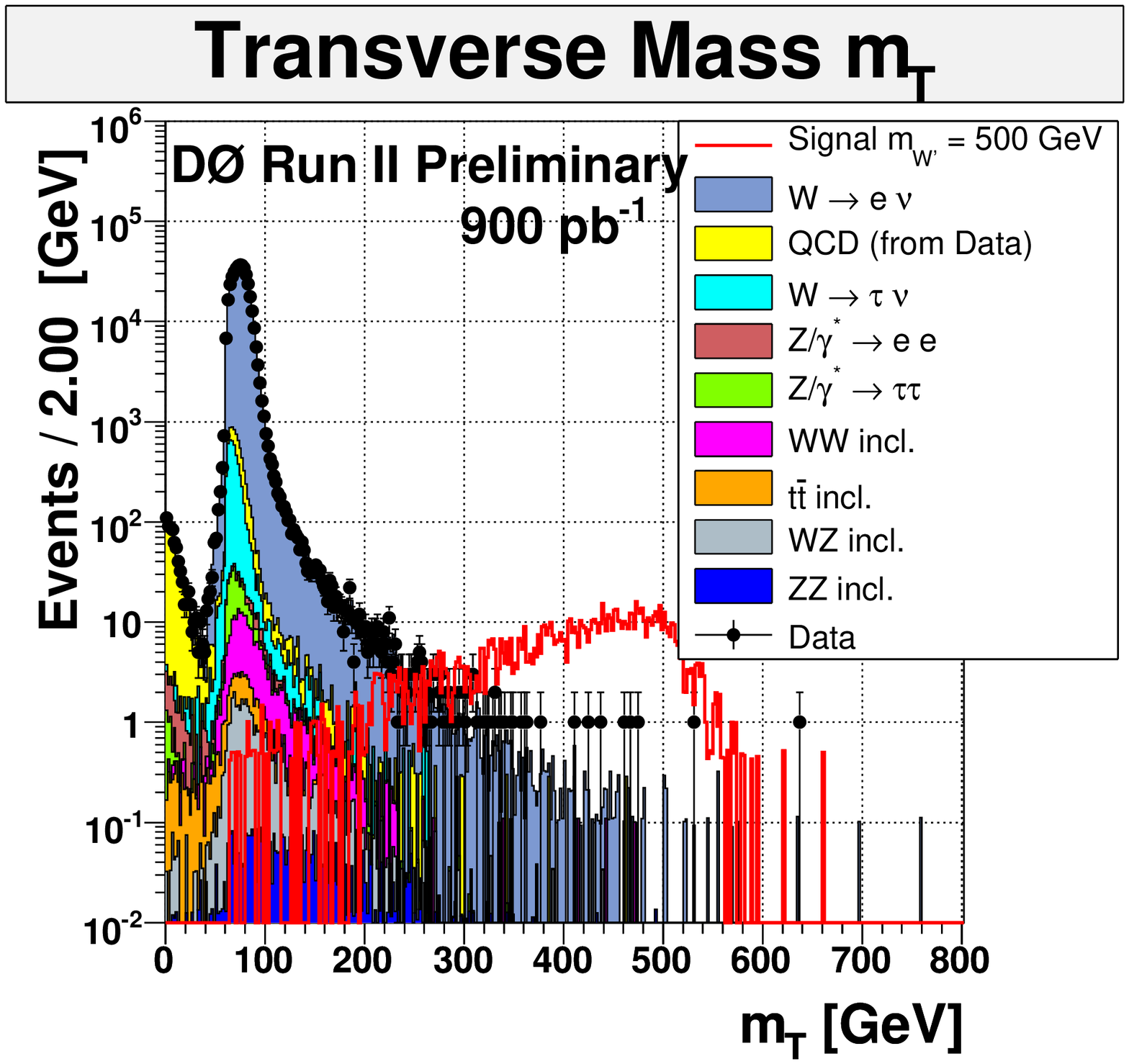} \hfill
\includegraphics[width=0.31\textwidth,angle=0]{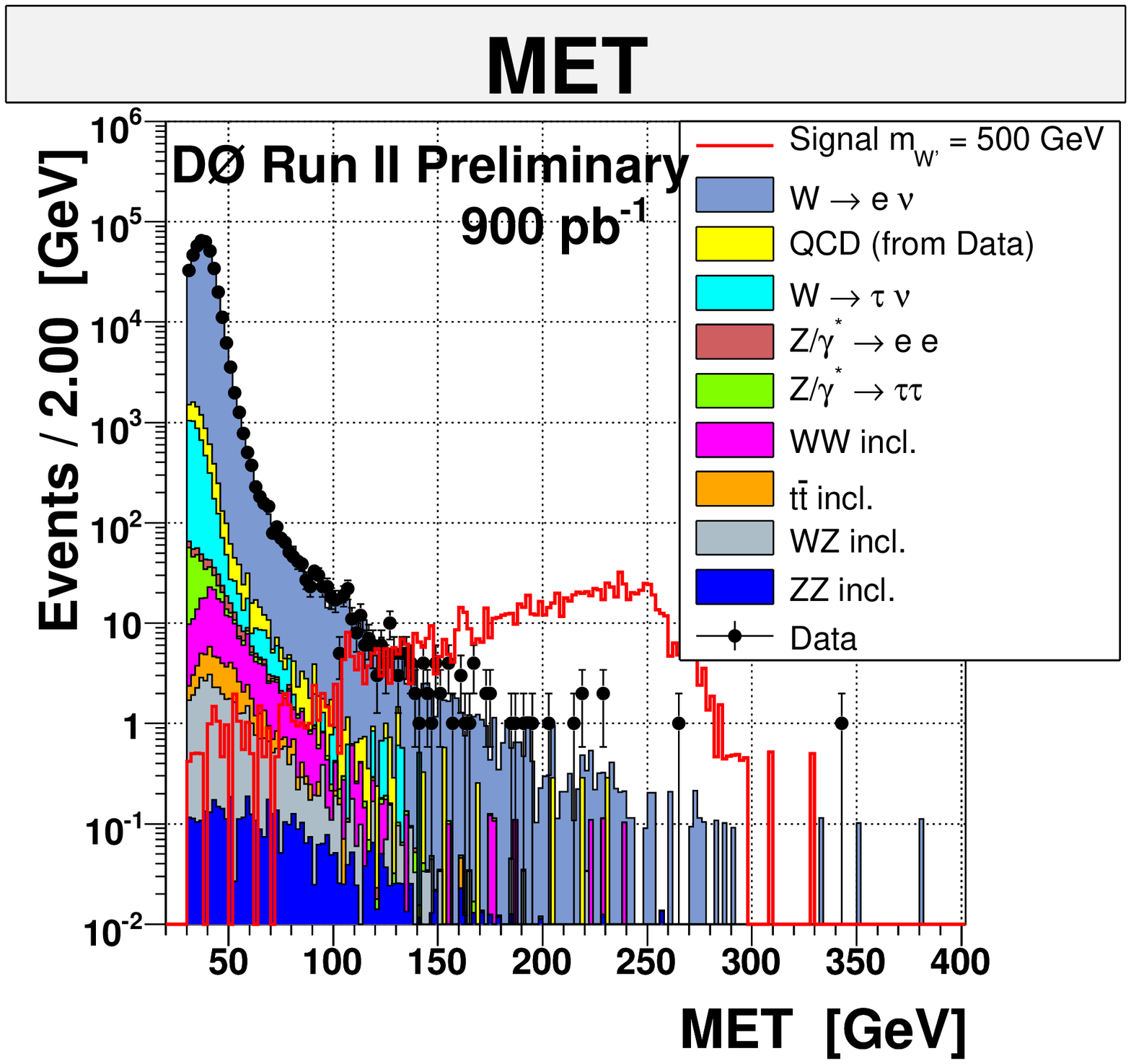} \hfill
\includegraphics[width=0.31\textwidth,angle=0]{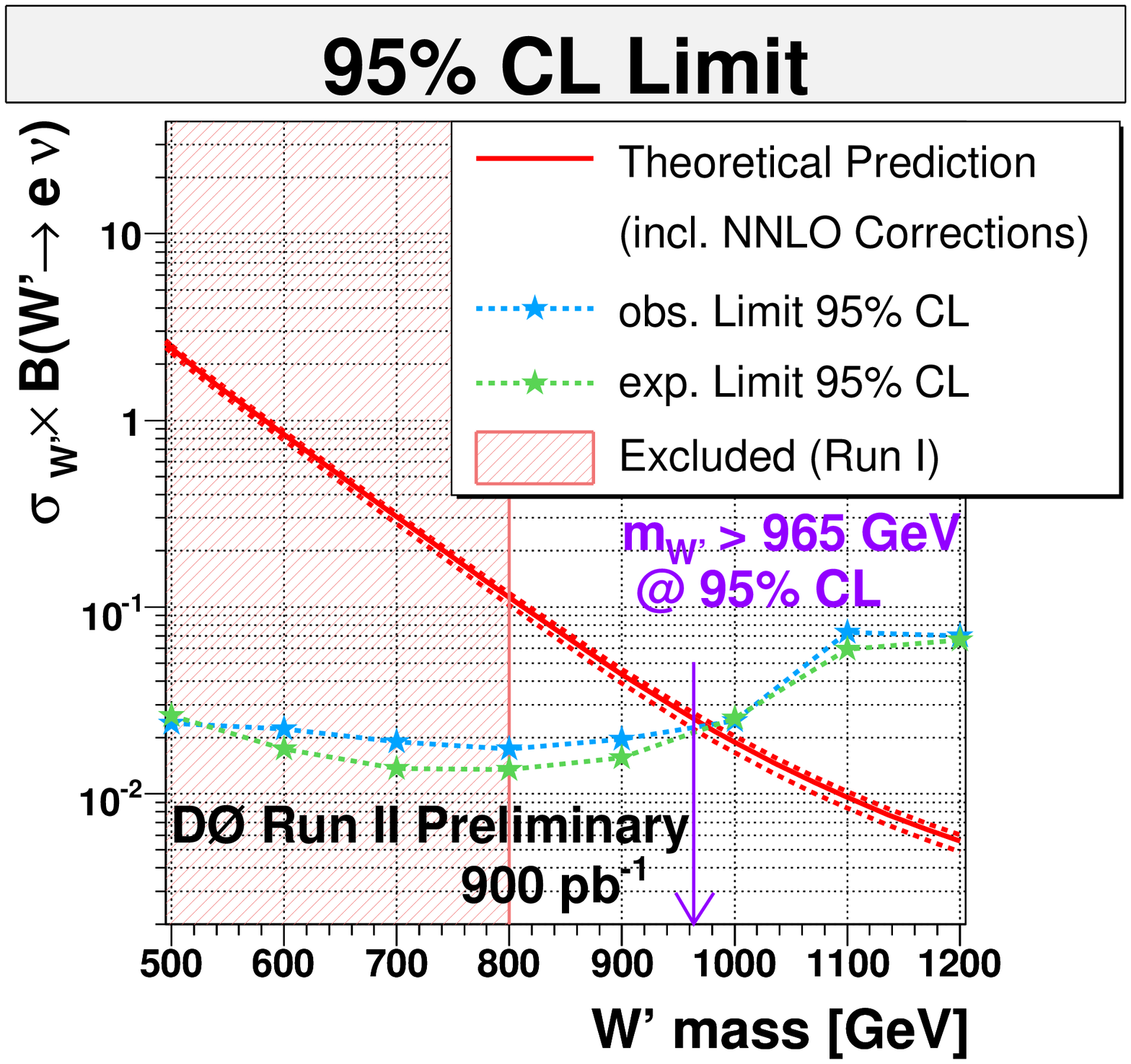}
\caption{Left: transverse mass $m_T$ distribution; middle: missing 
transverse energy $\MET$ distribution; right: 95\% CL upper limit on 
the production cross section times branching fraction
as a function of the $W'$ boson mass $m_{W'}$.}
\label{fig:WP}  
\end{figure*}

Additional charged gauge bosons $W'$ (as well as additional neutral gauge 
bosons $Z'$) are introduced in many extensions to the SM, e.\,g. 
Left-Right-Symmetric models (broken SU(2)$_L$ $\otimes$ SU(2)$_R$) or in 
GUT models which may also imply supersymmetry (e.\,g. $E_6$), see \cite{Moha}.
In some models the $W'$ boson is right-handed, and decays therefore into 
a right-handed neutrino and a charged lepton. However, such a neutrino has 
not yet been observed. Assuming the most general case, the new gauge group 
can comprise a new mixing angle $\xi$, new couplings to the fermions $g'$ 
and a new CKM matrix $U'$.  We make the assumptions that there is no 
mixing, $g'$ is equal to the SM coupling, $U'$ is equal to the SM CKM 
matrix, and that the new decay channel $W'\rightarrow WZ$ is 
suppressed \cite{Altarelli}. Furthermore, the width $\Gamma _{W'}$ of 
the $W'$ boson is assumed to scale with its mass $m_{W'}$. The decay into 
the third quark family (e.\,g. $W'\rightarrow t\bar b$), which is possible 
for $m_{W'}$ above 180~GeV, is taken into account. In case of the existence 
of additional generations of fermions, it is assumed that they are too 
heavy to be produced by a $W'$ decay. In this analysis the decay 
$W'\rightarrow e\nu $ is investigated.

Data events are selected if they are triggered, exhibit missing transverse 
energy $\MET>30$~GeV and contain an electron with transverse energy 
$E_T>30$~GeV within $|\eta|<1.1$. The electron is required to be isolated 
in the calorimeter and to have a track matched in $z$ and $\phi$ direction.
The energy deposition pattern of the electron is required to be consistent 
with EM showers. Further cleaning cuts (e.\,g. no jet activity in opposite 
direction of the electron or $\MET$) are applied in order to reject 
misidentified $\MET$ and the multijet background.

Several SM processes lead to the electron + $\MET$ final state, like 
$W\rightarrow e\nu$ or $W\rightarrow \tau\nu\rightarrow e\nu X$.
These backgrounds are estimated using \textsc{pythia} \cite{pythia} 
Monte Carlo samples. Events from multijet production can also contribute 
if one jet is misidentified as an electron and large $\MET$ is caused by 
energy mismeasurement. This contribution is estimated from a special data 
sample. In this sample the electron candidate fails the tight shower 
shape requirement. The resulting events are scaled to the data sample. 
The scale factor is adjusted in the low reconstructed mass region 
($m_T<30$~GeV, dominated by multijet events) to fill the missing events 
between the Monte Carlo prediction and observed data, see 
Fig. \ref{fig:WP} (left). The overall normalization is determined
in the $W$ peak region (60~GeV $<m_T<$ 140~GeV).

A good agreement between data and Monte Carlo prediction is observed, 
see Fig. \ref{fig:WP} (left and middle). In the data 630 events are 
reconstructed with $m_T>150$~GeV, compared to a background expectation 
of 623~$\pm$~81 events. Two kinds of systematic uncertainties contribute 
in this analysis: global normalization uncertainties (cross sections,
normalization, efficiency correction, multijet estimation) and shape changing 
uncertainties (electron energy scale and resolution, PDF uncertainty, 
uncertainty of the width of the $W$, jet energy scale). The overall systematic 
uncertainty on the background samples is of the order of 15\%; the uncertainty 
on the signal samples varies between 15\% and 50\%.

A Bayesian approach \cite{d0limit} with flat prior leads to upper limits on 
the production cross section times branching fraction, 
$\sigma(p\bar p\rightarrow W'+X) \times B(W'\rightarrow e\nu)$. The limits, 
shown in Fig. \ref{fig:WP} (right) as a function of the mass $m_{W'}$, 
are derived using a binned likelihood for the whole transverse mass 
spectrum $m_T>150$~GeV. Comparing the observed limit with the theoretical 
prediction additional heavy charged gauge bosons can be excluded at the 
95\% CL up to 965~GeV, hence significantly improving upon the previous best
limit of 800~GeV (D{\O} Run I \cite{oldWp}).

\section{Summary}

Final states with high energy objects are sensitive to a broad array of 
new physics. Three recent analyses have been presented which investigate 
the dielectron, diphoton, electron plus photon, and electron plus missing 
transverse energy final states using 1~fb$^{-1}$ of data taken with the 
D{\O} experiment at the Tevatron proton-antiproton collider. Since no 
significant excess is observed in the data, new restrictive limits 
are set in all three cases.

\end{document}